\documentclass[twocolumn]{aastex631}

\usepackage{xspace}
\usepackage{enumerate}
\usepackage{lineno}
\usepackage{array, booktabs, ragged2e}
\usepackage{float}

\newcommand{\yr}    	{\ifmmode \mathrm{yr} \else yr\fi}
\newcommand{\mpc}   	{\ifmmode \,\mathrm{Mpc}^{-3} \else \,Mpc$^{-3}$\fi}
\newcommand{\Msun}	    {\ifmmode \,\mathrm M_{\odot} \else $\,\mathrm M_{\odot}$\fi\xspace}
\newcommand{\Zsun}	    {\ifmmode \,\mathrm Z_{\odot} \else $\,\mathrm Z_{\odot}$\fi\xspace}
\newcommand{\Mhalo} 	{\ifmmode M_{\mathrm{halo}} \else $M_{\mathrm{halo}}$\fi\xspace}
\newcommand{\Rvir}  	{\ifmmode R_{200} \else $R_{200}$\fi\xspace}
\newcommand{\Mstar}	    {\ifmmode {M}_{\star} \else ${M}_{\star}$\fi\xspace}
\newcommand{\Mvir}	    {\ifmmode M_{\mathrm{halo}} \else $M_{\rm halo}$ \fi\xspace}
\newcommand{\Htwo}  	{\ifmmode {\rm H}_{2} \else ${\rm H}_{2}$ \fi}
\newcommand{\nH}    	{\ifmmode {n}_{\rm H} \else ${n}_{\rm H}$ \fi}
\newcommand{\cc}	    {\ifmmode {\rm cm}^{-3} \else ${\rm cm}^{-3}$ \fi}
\newcommand{\mperyr}	{\ifmmode \Msun{\rm yr}^{-1} \else $\Msun{\rm yr}^{-1}$ \fi}

\newcommand{\civ}	    {\ifmmode {\rm C}_{\rm IV} \else CIV \fi}
\newcommand{\mgii}	    {\ifmmode {\rm Mg}_{\rm II} \else MgII \fi}
\newcommand{\oi}	    {\ifmmode {\rm O}_{\rm I} \else OI \fi}
\newcommand{\perccm}	{\ifmmode {\rm cm}^{-2} \else ${\rm cm}^{-2}$ \fi}

\newcommand{\GD}[1]{{\color{black}{#1}}}

\shorttitle{SIDM in a Local Group Dwarf}
\shortauthors{Gutcke et al.}
\begin{document}

\title{Self-interacting dark matter in the center of a Local Group dwarf galaxy and its satellites}

\email{gutcke@hawaii.edu}
\author[0000-0001-6179-7701]{Thales A. Gutcke}
\affiliation{Institute for Astronomy, University of Hawaii, 2680 Woodlawn Drive, Honolulu, HI 96822, USA}
\author[0000-0001-6150-4112]{Giulia Despali}
\affiliation{Alma Mater Studiorum - Università di Bologna, Dipartimento di Fisica e Astronomia "Augusto Righi", Via Gobetti 93/2, I-40129 Bologna, Italy}
\affiliation{INAF-Osservatorio di Astrofisica e Scienza dello Spazio di Bologna, Via Gobetti 93/3, I-40129 Bologna, Italy}
\affiliation{INFN-Sezione di Bologna, Viale Berti Pichat 6/2, I-40127 Bologna, Italy} 
\author[0000-0002-7968-2088]{Stephanie O'Neil}
\affiliation{Department of Physics \& Astronomy, University of Pennsylvania, Philadelphia, PA 19104, USA}
\affiliation{Department of Physics, Princeton University, Princeton, NJ 08544, USA}
\author[0000-0001-8593-7692]{Mark Vogelsberger}
\affiliation{Department of Physics, Kavli Institute for Astrophysics and Space Research, Massachusetts Institute of Technology, Cambridge, MA 02139, USA}
\author[0000-0002-6831-5215]{Azadeh Fattahi}
\affiliation{The Oskar Klein Centre, Department of Physics, Stockholm University, Albanova University Center, 106 91 Stockholm, Sweden}
\author[0000-0002-1233-9998]{David B. Sanders}
\affiliation{Institute for Astronomy, University of Hawaii, 2680 Woodlawn Drive, Honolulu, HI 96822, USA}

\begin{abstract}

We present a detailed comparison of a Local Group dwarf galaxy analogue evolved in two cosmological models: the standard $\Lambda$CDM and a self-interacting dark matter (SIDM) model with a velocity-dependent cross-section. Both simulations are run with the high-resolution, hydrodynamical LYRA galaxy formation model, allowing us to explore the global and substructure properties of the dwarf in a consistent context. While the overall halo growth, final mass, and subhalo mass functions remain largely unchanged across models, SIDM produces a central dark matter core extending to $\sim$1 kpc, which does not significantly vary with the inclusion of baryons. Baryonic properties, however, differ notably. The SIDM model leads to a 25\% reduction in stellar mass and retains more gas within the stellar half-mass radius due to a prolonged quiescent phase in star formation. The stellar distribution is less centrally concentrated, and a population of in-situ star clusters form at late times. Substructure analysis reveals fewer luminous satellites and more stellar-only systems in SIDM, driven in part by tidal stripping that affects the dark matter more than the stars. A subset of satellites undergoes tidal-triggered core collapse after infall, enhancing the diversity of SIDM satellite rotation curves. These differences offer potential observational signatures of SIDM in low-mass galaxies.

\end{abstract}

\section{Introduction} \label{sec:intro}

Self-interacting dark matter (SIDM) represents a plausible class of alternative dark matter models, particularly in regimes where standard $\Lambda$ cold dark matter ($\Lambda$CDM) struggles to reproduce observed galaxy properties \citep[e.g.,][]{Spergel2000}. These tensions arise most notably at the extremes of the galaxy mass function: in the central regions of galaxy clusters \citep{Meneghetti2020,Meneghetti2023} and in the central regions of dwarf galaxies, particularly those orbiting the Milky Way (MW).

This work focuses on the latter case—Local Group (LG) dwarf galaxies. Observations of their dark matter distributions, often inferred from rotation curves, reveal a marked diversity of central density profiles. Some systems exhibit flat, cored profiles, while others show rising, cuspy behavior \citep{Oh2011, Adams2014, Kamada2017}. This diversity conflicts with $\Lambda$CDM predictions, which favor universally rising profiles shaped by the Navarro-Frenk-White (NFW) profile \citep{NFW}. However, studies have cautioned that observational methods, particularly those based on gas kinematics, may not accurately recover the true gravitational potential \citep[e.g.,][]{Oman2015}.

SIDM offers a physically motivated alternative that arises naturally in particle physics. In this framework, dark matter particles undergo non-gravitational scattering in dense environments, such as galaxy centers. These self-interactions redistribute kinetic energy, causing some particles to migrate outward and thereby lowering the central density to produce a core. Yet, core formation alone cannot explain the full observed diversity in dwarf galaxy profiles. Crucially, SIDM models predict that cores can become dynamically unstable under certain conditions, leading to core collapse—a process that can produce central densities even higher than those found in standard NFW cusps. Determining the conditions for this collapse and constraining the self-interaction cross-section remain key goals in current SIDM research.

Since SIDM was first proposed as a solution to small-scale structure problems by \citet{Spergel2000}, subsequent studies have explored its implications under various physical and numerical setups. \citet{Hannestad2000} investigated SIDM in the context of warm dark matter, while \citet{Burkert2000} used N-body simulations to make early predictions for SIDM halos. A major advance came with the implementation of a Monte Carlo treatment of elastic SIDM scattering in the Gadget3 \citep{Vogelsberger2012} and AREPO code by \citet{Vogelsberger2019}, which forms the basis for the present work. \GD{The AREPO SIDM implementation was successfully used at multiple scales, from dwarf galaxies, to MW-like galaxies \citep{Rose2023} and massive ellipticals \citep{Despali2019,Despali2022}}. This model was later extended to include inelastic interactions \citep{Vogelsberger2019}, and used by \citet{O'Neil2023} at the MW scale.
Reviews of SIDM and galaxy formation in this context can be found in \citet{Adhikari2022} and \citet{Vogelsberger2020}.

Zoom-in simulations have further constrained SIDM models. \citet{Zavala2013} showed that a constant self-interaction cross-section cannot simultaneously explain both MW satellite rotation curves and galaxy cluster observations. Cross-sections larger than $\sigma/m_{\chi} \gtrsim 1~\text{cm}^2/\text{g}$ tend to overproduce cores in clusters, contradicting gravitational lensing constraints \citep{Robertson2019}. These findings point toward a velocity-dependent cross-section as a necessary feature of viable SIDM models. However, \cite{Errani2023} show that cores larger than 1\% of their scale radius in Milky Way subhalos would greatly increase their disruption. This makes the ultra-faint satellites Tuc 3, Seg 1, Seg 2, Ret 2, Tri 2, and Wil 1 incompatible with an SIDM cross-section that increases their core sizes.

Recent work has focused on modeling the velocity dependence and its implications. The TangoSIDM project \citep{Correa2022, Correa2025}, the suite of simulations by \citet{Shah2024} \GD{and the cosmological boxes of the AIDA-TNG project \citep{Despali2025}} explore a wide range of host halo masses ($10^{11}$–$10^{15}\Msun$) and cross-section scalings. 

Several studies have investigated the mechanisms and conditions behind SIDM core collapse. \citet{Elbert2015} and \citet{Nishikawa2020} demonstrate that tidal stripping in subhalos can accelerate collapse. \citet{Shah2024} show that subhalos with masses $\lesssim 10^9\Msun$ can undergo core collapse within a Hubble time when exposed to strong tidal fields, with the collapse fraction sensitively dependent on the low-velocity behavior of the cross-section. 
Using dark matter-only simulations in a static host potential, \citet{Zeng2022} find that collapse is strongly influenced by halo concentration and that subhalo evaporation delays the process. They argue that constant cross-section SIDM is unlikely to produce widespread collapse unless baryonic effects are included. Indeed, \citet{Nadler2023} simulate group-scale ($10^{13}~\Msun$) halos with velocity-dependent SIDM, finding that dense subhalos—ideal core-collapse candidates—can emerge naturally and match the properties of perturbers inferred from strong lensing.

\citet{Turner2021} use MW-analogue simulations to show that SIDM core collapse can explain the diverse central densities observed in MW satellites, particularly in subhalos with masses $5 \times 10^6$–$10^8~\Msun$. Similarly, \citet{Sameie2018} show that baryonic potentials (modeled as static disks) enhance core collapse, and that stronger self-interactions lead to earlier and more dramatic collapse. \citet{Despali2019}, using zoom-in hydrodynamic simulations, find that SIDM halos in the MW mass range can develop cuspier profiles than their CDM counterparts, while early-type galaxy analogues remain cored.

Despite these advances, many past studies have limitations. Most are based on dark matter-only simulations, sometimes supplemented with semi-analytic or static baryonic potentials, which lack live gas–dark matter interactions. This is a significant drawback, as observational inferences of dark matter profiles often rely on gas dynamics. Moreover, simulations that include baryons are often non-cosmological, missing realistic halo assembly histories.

To overcome these limitations, we present fully cosmological, hydrodynamic zoom-in simulations with live dark matter and baryons. We focus on a low-mass dwarf galaxy and its substructure, motivated by the growing potential of substructure lensing to detect halos with masses below $10^{10}~\Msun$.

The paper is organized as follows. Section~\ref{sec:model} introduces our simulation framework and SIDM parameter settings. In Section~\ref{sec:results}, we present our findings. We begin by analyzing the properties of the main dwarf halo, including its dark matter profile (Sec.~\ref{sec:DMprofiles}), baryonic response (Sec.\ref{sec:baryons}), and star formation history (Sec.~\ref{sec:sfh}). We then examine the galaxy’s substructure (Sec.~\ref{sec:subhalos}) and investigate evidence for core collapse (Sec.~\ref{sec:corecollapse}). Our conclusions are summarized in Section~\ref{sec:conclusion}.

\section{Model \& Method} \label{sec:model}

\begin{figure}
    \centering
    \includegraphics[width=\columnwidth]{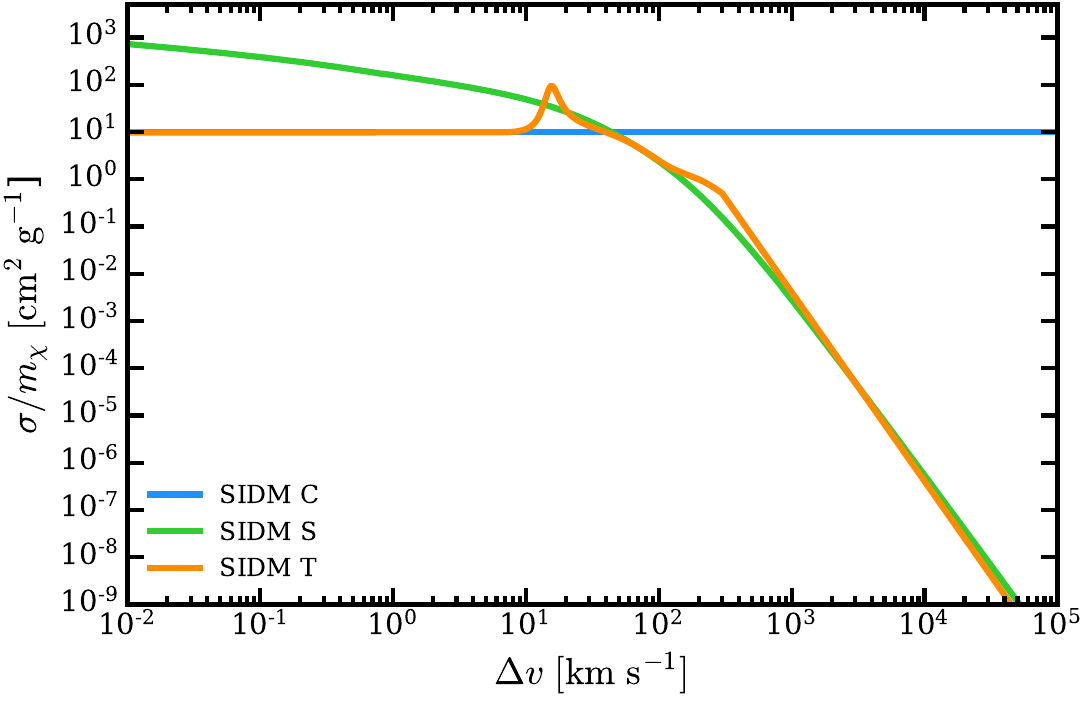}
    \includegraphics[width=\columnwidth]{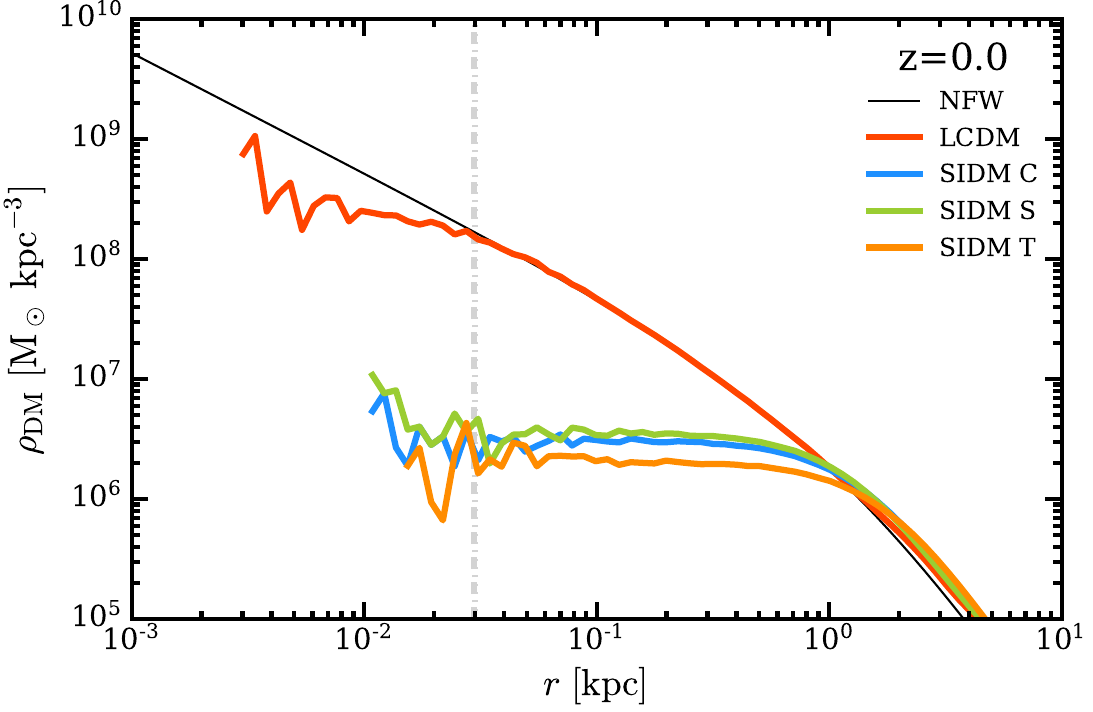}
    \caption{Three SIDM models with differing velocity dependency. \GD{The top panel shows the cross-section as a function of the relative velocity of particles, which is constant for the "SIDM C" (light blue) model and instead has a steep dependence on velocity in "SIDM S" (green) and "SIDM T" (orange). The bottom panel shows the dark matter density profile at $z=0$ of the halo in the dark-matter-only run.} In this work, we will study the effects of the "SIDM S" (green) model and its interaction with baryonic physics in more detail.}
    \label{fig:models}
  \end{figure}

In this study, we employ the LYRA cosmological, hydrodynamical galaxy formation model \citep{Gutcke2021, Gutcke2022a}, which has been shown to produce realistic analogues of Local Group dwarf galaxies at $z=0$ \citep{Gutcke2022b}. In particular, these simulations reproduce stellar properties at $z=0$, including stellar mass, size, kinematics, metallicity, and star formation history. For the purposes of this work, we re-simulate the galaxy HaloF presented in \citet{Gutcke2024} using self-interacting dark matter (SIDM) models and compare it to its original $\Lambda$CDM counterpart. This galaxy has a virial mass of $9 \times 10^9~\Msun$ and a corresponding stellar mass of $10^7~\Msun$ at $z=0$. It continues to form stars at low rates—between $10^{-4}$ and $10^{-6}~\mperyr$—throughout cosmic time, until the simulation ends at $z=0$.

The LYRA model is a comprehensive numerical framework that incorporates a resolved interstellar medium (ISM) with a cooling prescription operative down to temperatures as low as 10 K. It further includes individual, star-by-star star formation, resolved supernova events, and a subgrid model accounting for Population III (PopIII) star enrichment during the high-redshift epoch. These prescriptions are implemented within the cosmological, hydrodynamical moving-mesh code \textsc{Arepo} \citep{Springel2010, Pakmor2016, Weinberger2020}. For a detailed description of the model and its implementation, we refer the reader to the cited works. In the following, we highlight only the aspects most pertinent to this study.

To create the initial conditions, halos are selected at $z=0$ from the EAGLE simulation box, following the methodology of \citet{Jenkins2013}. Particles within a multiple of the halo’s virial radius are then traced back to their positions at $z=128$. The region encompassing these particles defines the Lagrangian region, effectively the accretion region of the halo. The full $L=100$cMpc box is retained at low resolution, with a high-resolution region embedded at the location of the Lagrangian volume. Within this region, the gas mass resolution is set to $4\Msun$, and the dark matter particle mass to approximately $80~\Msun$, significantly higher than in the surrounding low-resolution volume. To maintain consistent mass resolution during the simulation, gas cells are allowed to refine and de-refine adaptively. The gravitational softening lengths are 4pc for gas and stars, and 10pc for dark matter. The selection of target halos also follows an isolation criterion, ensuring they do not experience interactions with more massive galaxies over their lifetimes.

For the SIDM simulations, we use the model implemented by \citet{Vogelsberger2012} and \citet{Zavala2013}, where elastic scatterings between dark matter particles are mediated by an attractive Yukawa potential. SIDM simulations start from the same initial conditions as the CDM counterparts, as self-interactions do not affect the initial linear perturbations and only become important during the late, non-linear stages of structure formation. A Monte Carlo approach is used to model the dark matter self-interactions: particle pairs are randomly selected for scattering and are assigned new velocities with isotropically drawn directions. The timestep is chosen to be sufficiently small to avoid multiple scatterings within a single step \citep[see also][]{Vogelsberger2019}.

As shown in Fig.~\ref{fig:models}, we first consider three SIDM models in a dark matter-only set up. The top panel sows the velocity dependence of the cross-section and the lower panel shows the resulting DM density profile. Each characterized by a different velocity dependence of the scattering cross-section, and we show that they all produce a similar size core.
One model has a constant cross-section, $\sigma/m_{\chi} = 10~\mathrm{cm}^2/\mathrm{g}$ (SIDM C). The other two are velocity-dependent models: one based on the empirical fit by \citet{Correa21}, which was calibrated on observations of Milky Way dwarf spheroidal galaxies (SIDM S) and a second velocity-dependent model that includes a resonance at $v \sim 10$~km/s, enhancing the cross-section at the characteristic velocities of dwarf galaxies (SIDM T).

Because our focus is on SIDM effects in the low-mass, low-velocity regime typical of dwarf galaxies and their satellites, we select models that feature large cross-sections in this range. Velocity-dependent cross-sections are particularly promising because they reconcile the high cross-sections favored on small scales with the stringent upper limits derived from galaxy clusters and massive galaxies. The phenomenon of resonances in the velocity dependency (such as the one seen in "SIDM T" around 10-20 $km s^{-1}$) are emerging as interesting avenue to enhance cross sections locally \citep{Tran2025}.

While all three models are run with gravity only, for the hydrodynamical simulation in this work, we adopt the “SIDM S” model, which has the largest cross-section at dwarf galaxy scales. This makes it the most likely to produce core-collapsed structures at the masses probed in our study. Furthermore, as it is empirically calibrated on observations, it represents the most physically motivated scenario among the models considered. \GD{For simplicity, we refer to this model simply as SIDM in the rest of the paper.}

\section{Results} \label{sec:results}

\subsection{Dark matter density profiles}
\label{sec:DMprofiles}

\begin{figure}
    \centering
    \includegraphics[width=\columnwidth]{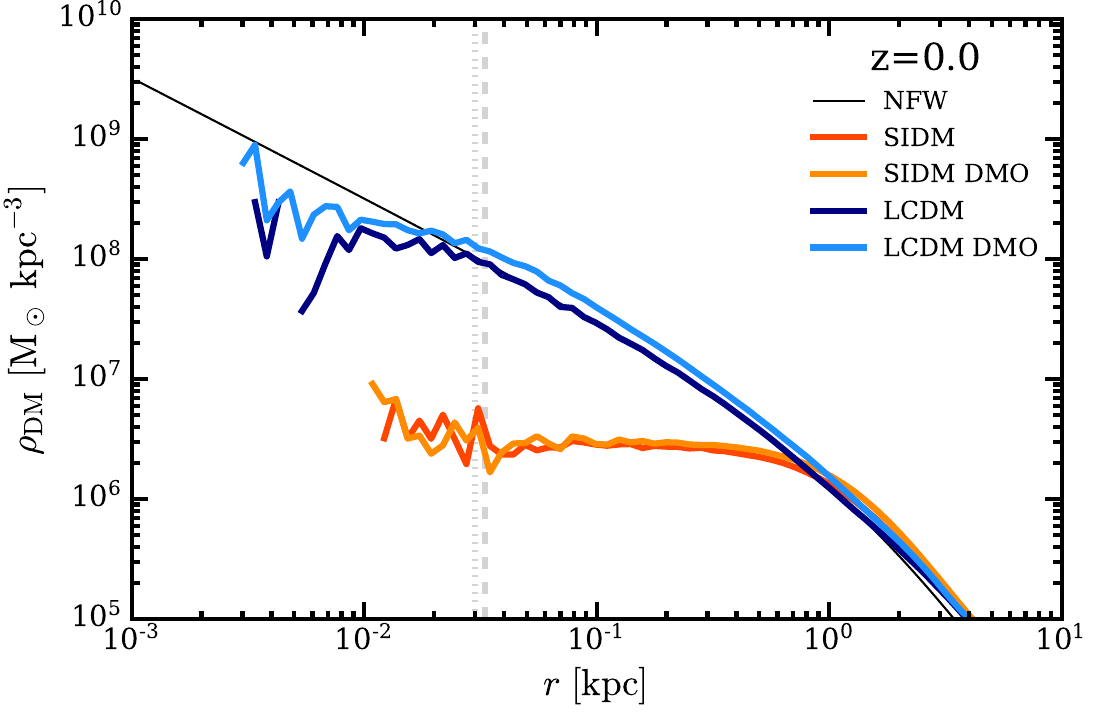}
    \includegraphics[width=0.95\columnwidth]{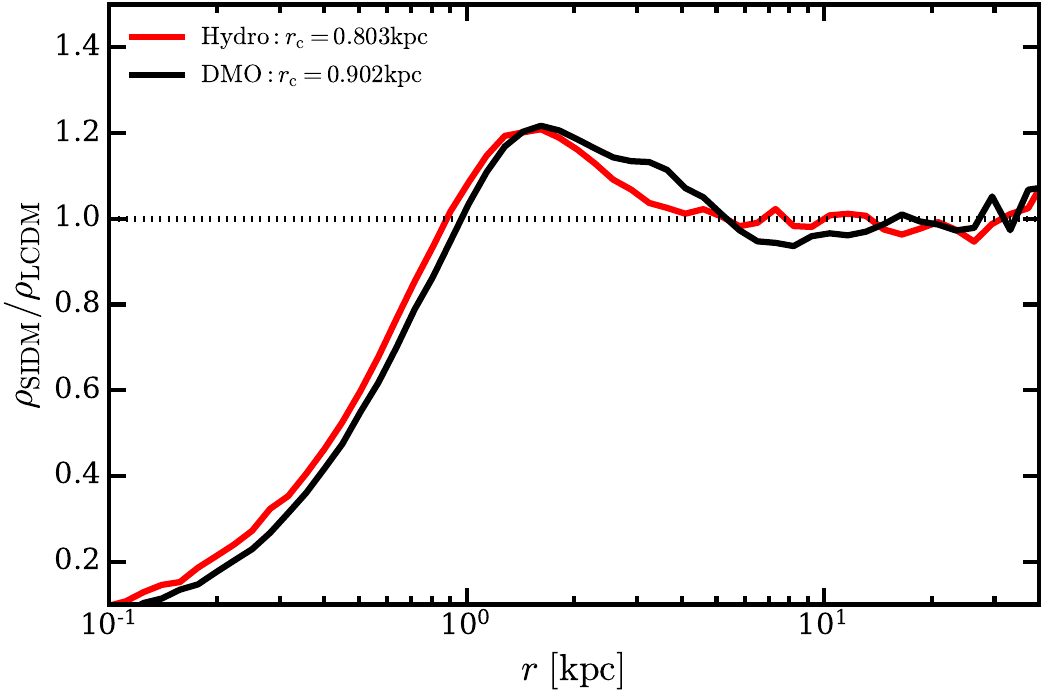}
    \includegraphics[width=\columnwidth]{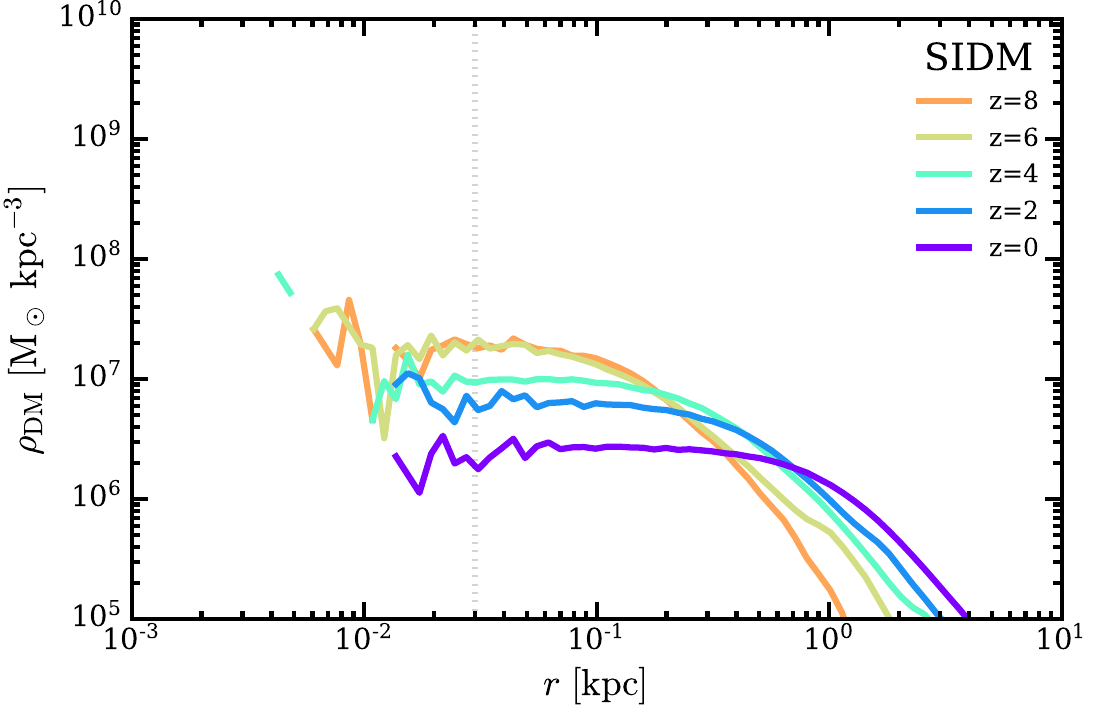}
    \caption{\textit{Top:} Dark matter density profiles. For the DMO simulations, we have reduced the mass by $(1-\Omega_\mathrm{B})$. The SIDM model produces a uniform density region at the center (``core'') out to a radius of approximately $1\mathrm{kpc}$. The grey solid line designates 3 times the DM softening length. The dashed gray line encloses 2000 DM particles. 
    \textit{Middle:} SIDM profile divided by $\Lambda$CDM profile for hydrodynamical and dark matter-only simulations. Core size as shown in legend is defined as the profile ratio reach $90\%$. It is a common SIDM prediction to see a ``bump" in the relative profiles where the excess DM that has been pushed out of the core piles up.
    \textit{Bottom:} The evolution of the SIDM profile over time. The core is formed very early, before $z=8$, and grows in size as the halo grows with time.}
    \label{fig:rhoDM}
\end{figure}

We begin by examining the impact of SIDM on the dark matter distribution at $z=0$ for our Local Group dwarf galaxy. It is well-known that self-interactions can produce a central core in dark matter haloes of any mass, provided that the cross-section is efficient at that scale. This, in turn, can affect the halo evolution and the properties of the baryonic component.
In the top panel of Fig.~\ref{fig:rhoDM} we present the dark matter density profile at $z=0$ of the four main simulations, namely $\Lambda$CDM, $\Lambda$CDM DMO, SIDM and SIDM DMO. The SIDM models are run with a velocity-dependent cross-section, shown as the green line in Fig.~\ref{fig:models}. The CDM simulations are presented in shades of blue, with the dark blue being the hydrodynamical simulation. The SIDM results are shown in shades of red, with the DMO simulation in dark orange. The central dwarf galaxy follows an NFW-like profile in the CDM runs. The thin black line is the NFW fit to the $\Lambda$CDM profile. In contrast, the SIDM runs produce a well-defined core with a radius of approximately 1 kpc. Notably, the presence of baryons does not significantly alter the core size or shape in this system. We see this in the central panel, where we show the ratio between the SIDM and the $\Lambda$CDM model for full run (in red) and dark matter-only version (in black). What we see here is the deficit of dark matter in the central region out to approximately 1 kpc. The excess dark matter between 1 and 3 kpc is a well-known feature in SIDM models, resulting from scattered particles accumulating at the core's edge.

\begin{table}[]
    \centering
    \begin{tabular}{llrr}
        & Unit & SIDM & $\Lambda$CDM \\
        \hline \hline
    Stellar mass & \Msun &  $1.86\times10^6$  & $2.48\times10^6$ \\
    Stellar half mass radius & kpc &  1.57  & 0.891 \\
    HI mass & \Msun &  $1.52\times10^6$  & $5.27\times10^5$ \\
    HI half mass radius & kpc &  0.705  & 0.243 \\
    \hline \hline
    \end{tabular}
    \caption{Simulated galactic values}
    \label{tab:general}
\end{table}

The bottom panel clearly shows the core formation begins at very early times. The colored lines show the dark matter density profile at different redshifts, as indicated in the legend. We see a flat central region as early as $z=8$. As the galaxy and its halo grow significantly in mass from $z=8$ to $z=0$, the radial extent of the cored region increases accordingly, from ~200 pc to ~1 kpc.

\subsection{Baryonic component}
\label{sec:baryons}

An important question in the context of SIDM is how the altered dark matter distribution influences the host galaxy and its evolution, particularly its baryonic component. This is especially true for dwarf galaxies since one of the aims of introducing SIDM is to better match the diversity of rotation curve shapes and central densities in this regime. The chosen SIDM cross-section is expected to produce a strong effect. As we will see in the following, the gas density distribution and morphology clearly show an alteration at $z=0$. 

\begin{figure}
    \centering
\includegraphics[width=\columnwidth]{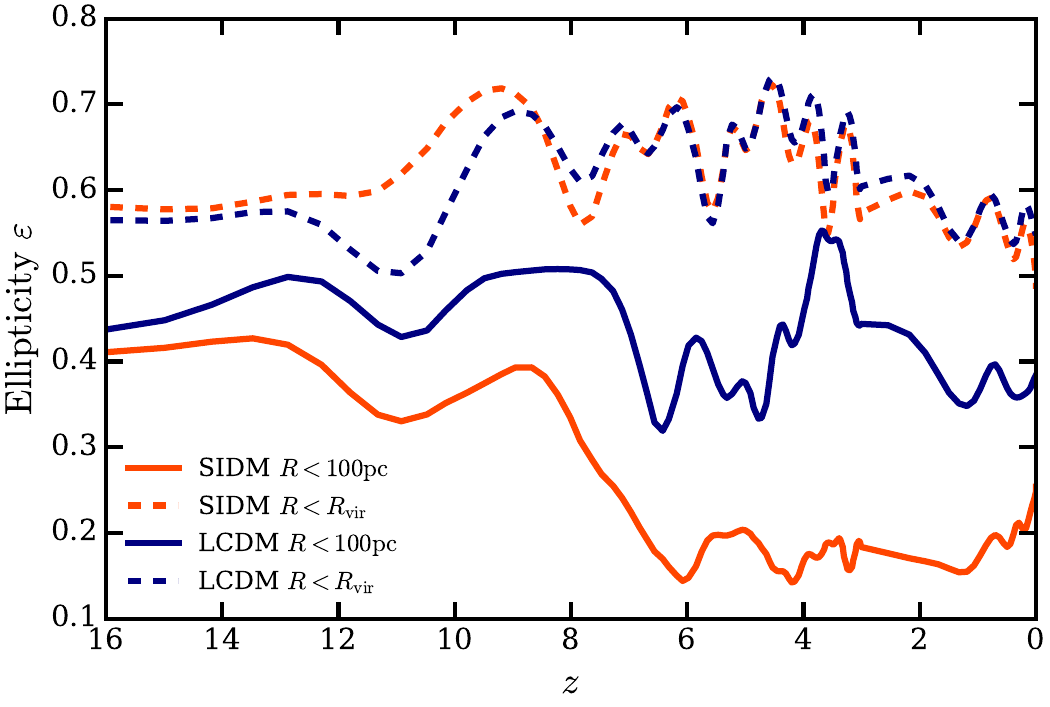}
    \caption{Ellipticity of the DM measured within $100~\mathrm{pc}$ and within \Rvir. }
    \label{fig:ellipticity}
\end{figure}

\begin{figure*}
    \centering
    \includegraphics[width=\textwidth]{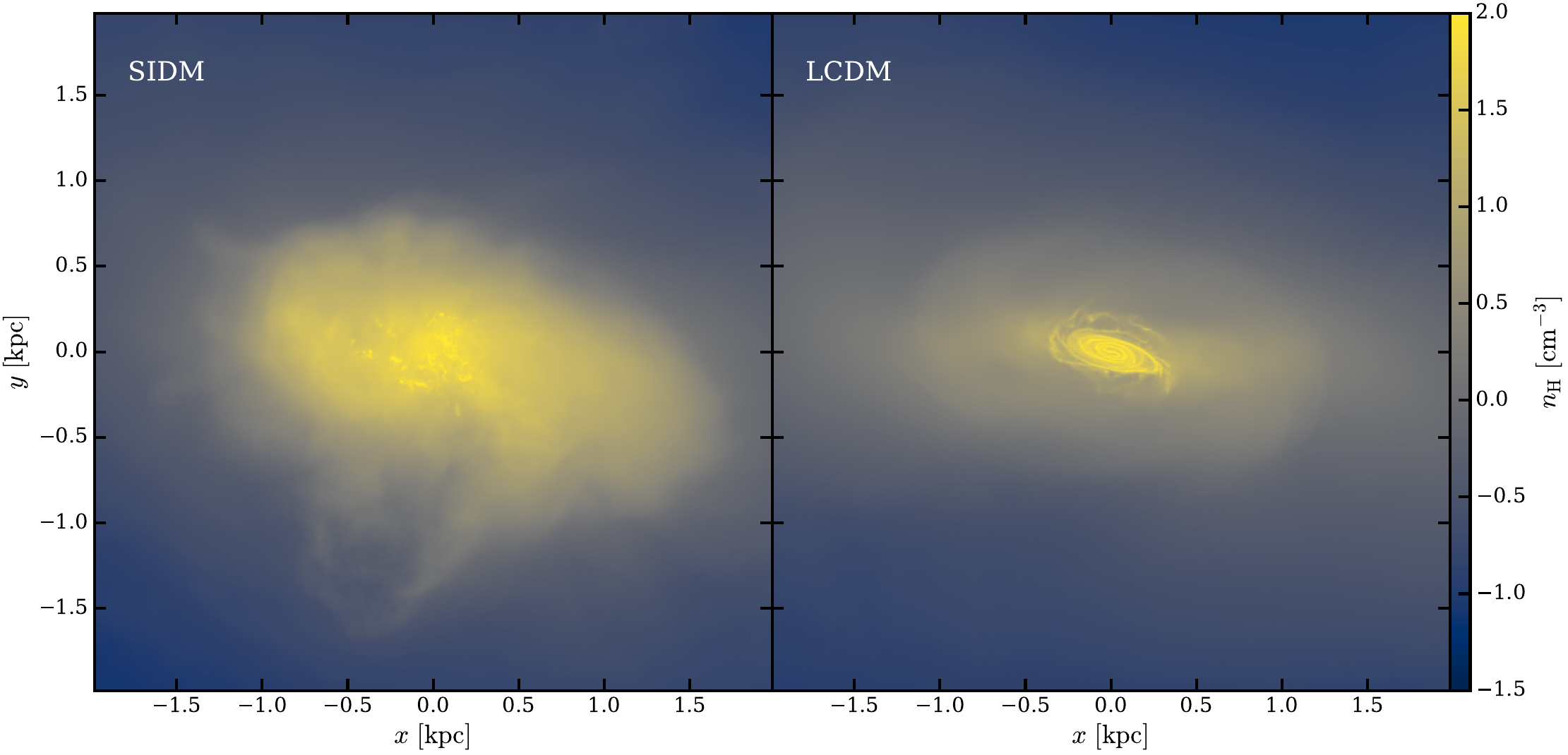}
    \includegraphics[width=\textwidth]{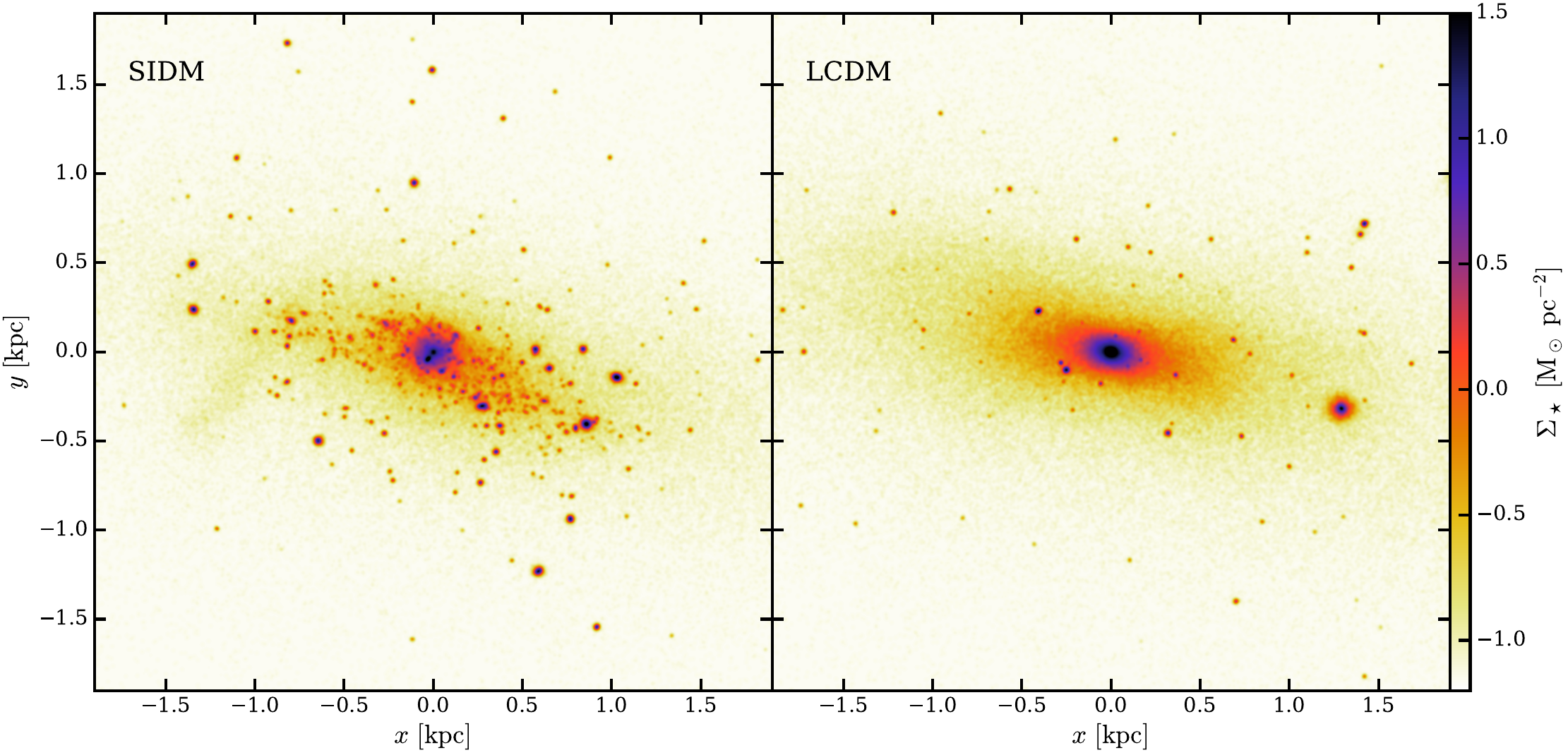}
    \caption{\textit{Top:} Neutral hydrogen density projection at $z=0$. Due to the larger core ($\sim 1\mathrm{kpc}$) in the SIDM run, the dense, star forming gas is more extended. \textit{Bottom:} Stellar surface density in the central 4 kpc. Due to the larger core, the SIDM run has less centrally concentrated SF, instead displaying more, smaller star formation regions, extending out to around $1\mathrm{kpc}$, the approximate size of the DM core.}
    \label{fig:barymap}
\end{figure*}

In Fig.~\ref{fig:barymap} we show the neutral hydrogen density for SIDM and $\Lambda$CDM in the top panels. In the lower panels are the stellar surface densities for the two simulations. As we can see, the SIDM galaxy has a more extended gas distribution, out to approximately the DM core radius of 1~kpc. The $\Lambda$CDM galaxy has a much smaller, more organized gas disk concentrated in the central few hundred parsec. This altered dense gas morphology gives rise to a different stellar morphology, too. In $\Lambda$CDM, most stars are concentrated in a single, dense central spheroid. In contrast, the SIDM core enables multiple star formation sites, resulting in a broader stellar distribution composed of distinct star clusters.

As we show in Fig.~\ref{fig:ellipticity}, the ellipticity of the DM distribution on the scale of the whole halo is equally elongated in both SIDM and $\Lambda$CDM. However, measured in the inner $100~\mathrm{pc}$ the SIDM halo shows a markedly stronger signature of sphericity. This is due to the fact that the baryons never dominate the potential in the SIDM case, whereas in $\Lambda$CDM the DM follows the gas in the central region. The ellipticity is measured by first computing the inertia tensor of the DM particles and then using the eigenvalues to obtain the principal axes.

\begin{figure}
    \centering
    \includegraphics[width=\columnwidth]{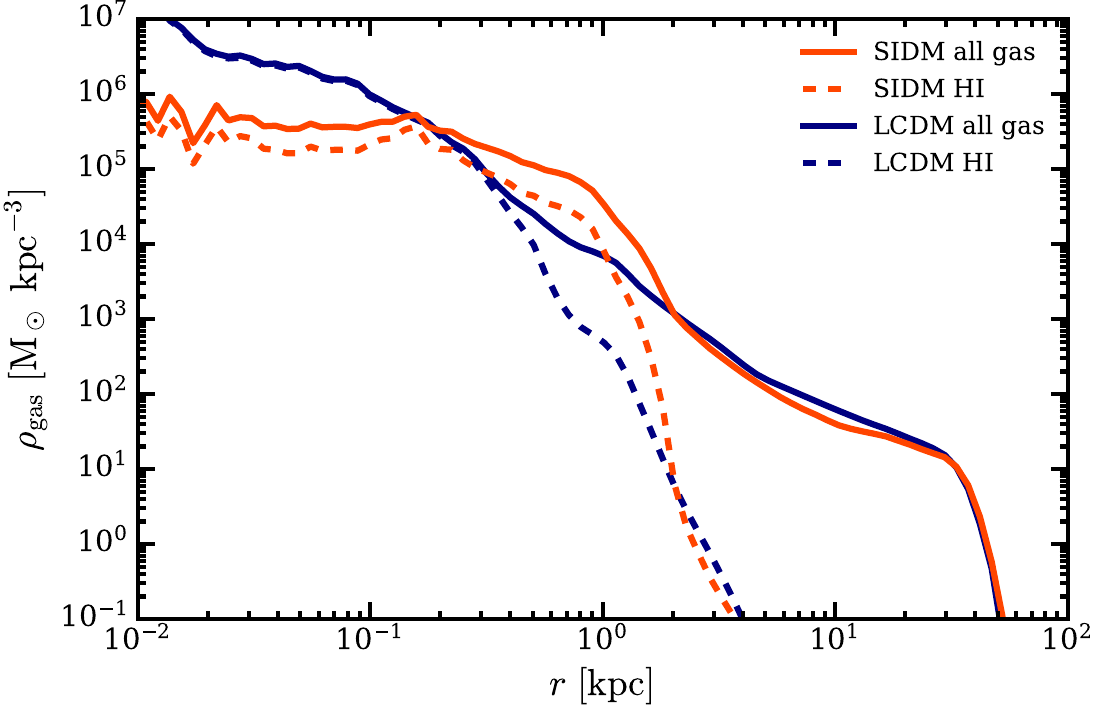}
    \includegraphics[width=\columnwidth]{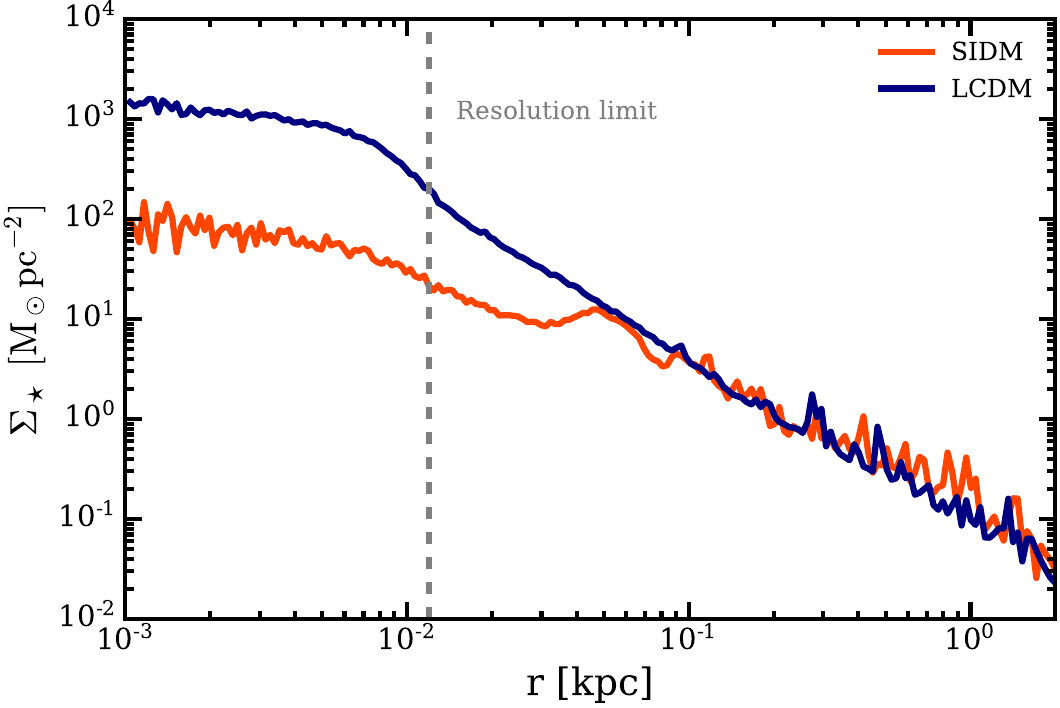}
    \caption{Top: Gas density and HI density profiles. Bottom: Stellar surface density profile. Due to the DM core, the SIDM run has less centrally concentrated SF, instead displaying more numerous and smaller star formation regions. The lower stellar surface density extends out to around $50\mathrm{pc}$, only about 5\% of the size of the DM core.}
    \label{fig:starsigma}
\end{figure}

Finally, in Fig.~\ref{fig:starsigma}, we show the gas density  and stellar surface density profiles for both SIDM and $\Lambda$CDM galaxies at $z=0$. The surface density in the outskirts (beyond 100~pc) remains very similar. The local maxima (spikes in the stellar profile) indicate the presence of luminous satellites. However, as we look toward the center, the stellar profile of the SIDM galaxy begins to depart from the $\Lambda$CDM profile at around 10\% of the DM core radius. The result is a significantly reduced stellar surface density in the inner regions, mirrored in the distinct stellar and HI half mass radii (see Tab.~\ref{tab:general}). In the lower panel, the gray dashed line marks 12~pc, equivalent to three times the stellar softening length, a measure of the resolution of the simulation. At radii smaller than this value numerical effects from the softening of the gravitational forces begin to dominate the profile, making it unreliable for astrophysical prediction. However, the departure of the two profiles appear well outside of this region, showing that there is an observable difference in central surface densities between the two models.

\begin{figure*}
    \centering
    \includegraphics[width=\columnwidth]{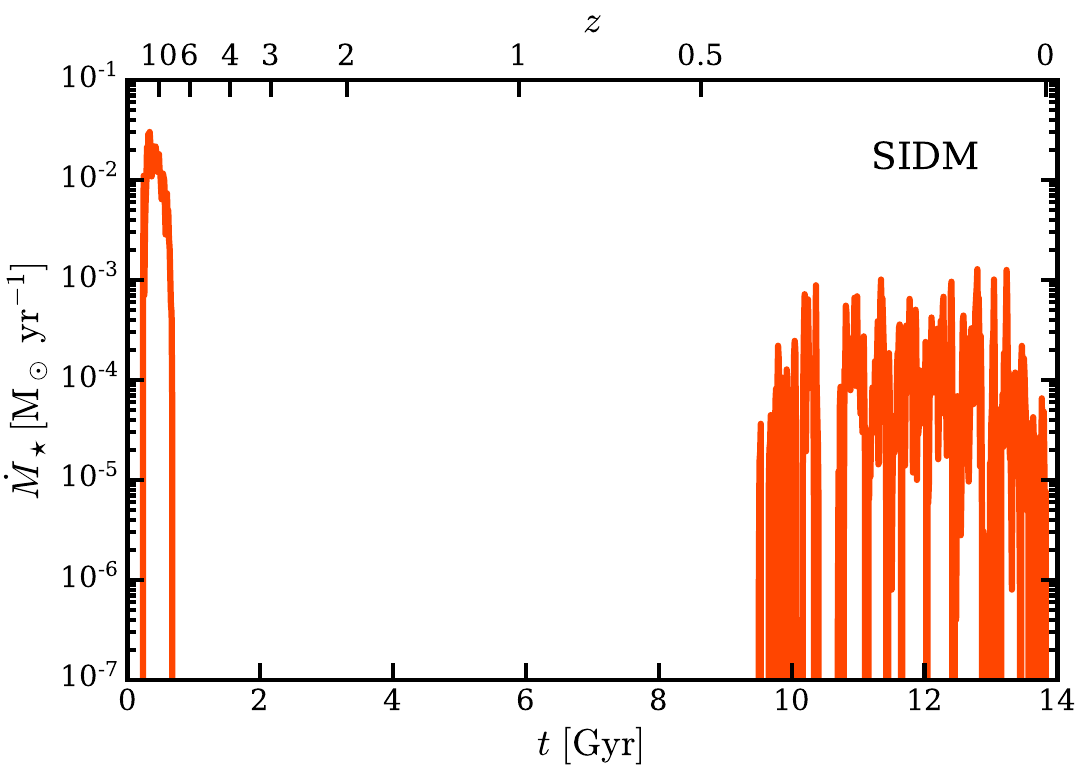}
    \includegraphics[width=\columnwidth]{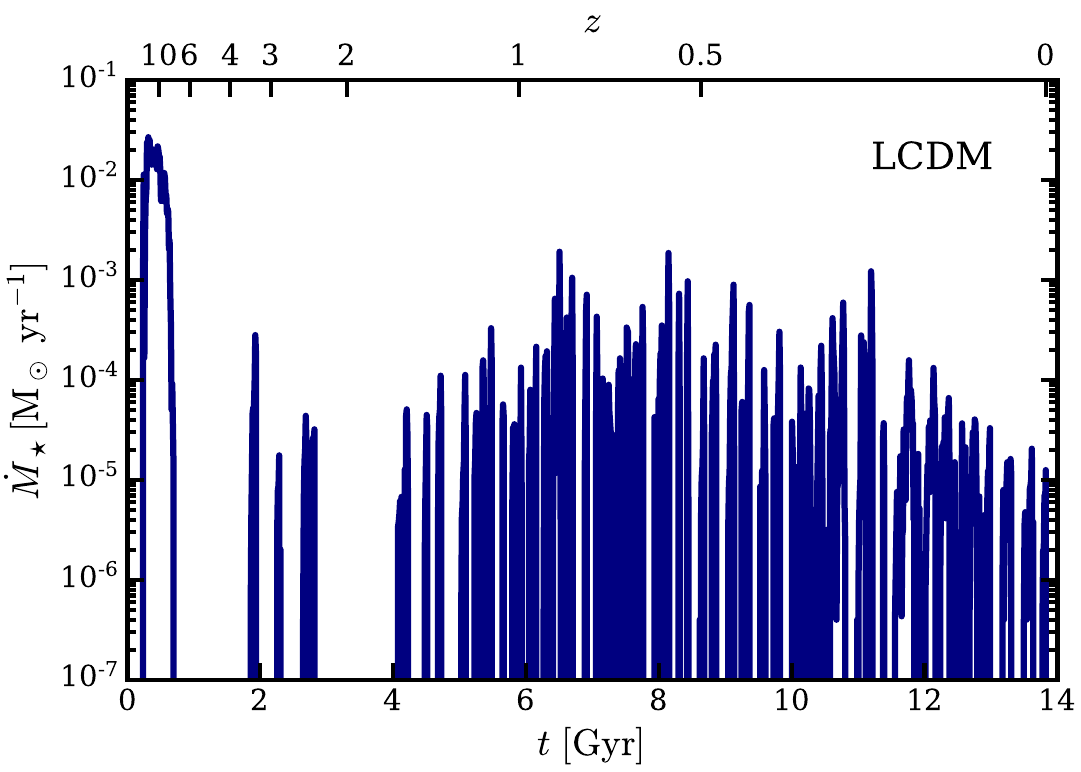}
    \caption{Star formation history for the SIDM and $\Lambda$CDM simulations. Both simulations have an initial burst of star formation before reionization, which is quenched due to the increased UV background. The $\Lambda$CDM run restarts SF more quickly, since the steeper DM profile funnels gas to the center more efficiently. By $z=0$, there is a still a 25\% difference in the total stellar mass formed: $1.9\times10^6\Msun$ and $2.5\times10^6\Msun$ for SIDM and $\Lambda$CDM, respectively.}
    \label{fig:sfh}
\end{figure*}

\subsection{Star formation history}
\label{sec:sfh}

SIDM not only alters the stellar distribution at $z=0$, but also impacts the galaxy’s star formation (SF) history over cosmic time. The $\Lambda$CDM galaxy (Fig.~\ref{fig:sfh}, right) forms a large amount of stars before the onset of Reionization (occurring at $\bar{z}=7.5$). The external ultra-violet background radiation causes the SF to be halted. However, the gradual re-accretion of gas allows sufficiently dense gas to accumulate, until it is able to shield itself from the radiation. This triggers the formation of cold, star forming gas and a re-onset of SF at around $z=5$ \citep[see][for a detailed analysis]{Gutcke2022b}. SF is then maintained in an extremely bursty state and at low rates for the remaining Cosmic time.

This evolution is altered in the SIDM universe. Due to the early formation of the DM core well before the onset of Reionization, the re-accretion of gas and the self-shielding is delayed by many billions of years (Fig.~\ref{fig:sfh}, left). In the SIDM simulation, star formation does not resume until $z\sim0.3$. Interestingly, despite the delayed onset, the total stellar mass differs by only ~25\%, suggesting that the prolonged gas accumulation in the SIDM run is sufficient to nearly catch up to $\Lambda$CDM by $z=0$. We can see this in the slightly higher SF rates at late times.

\begin{figure*}
    \centering
    \includegraphics[width=0.49\textwidth]{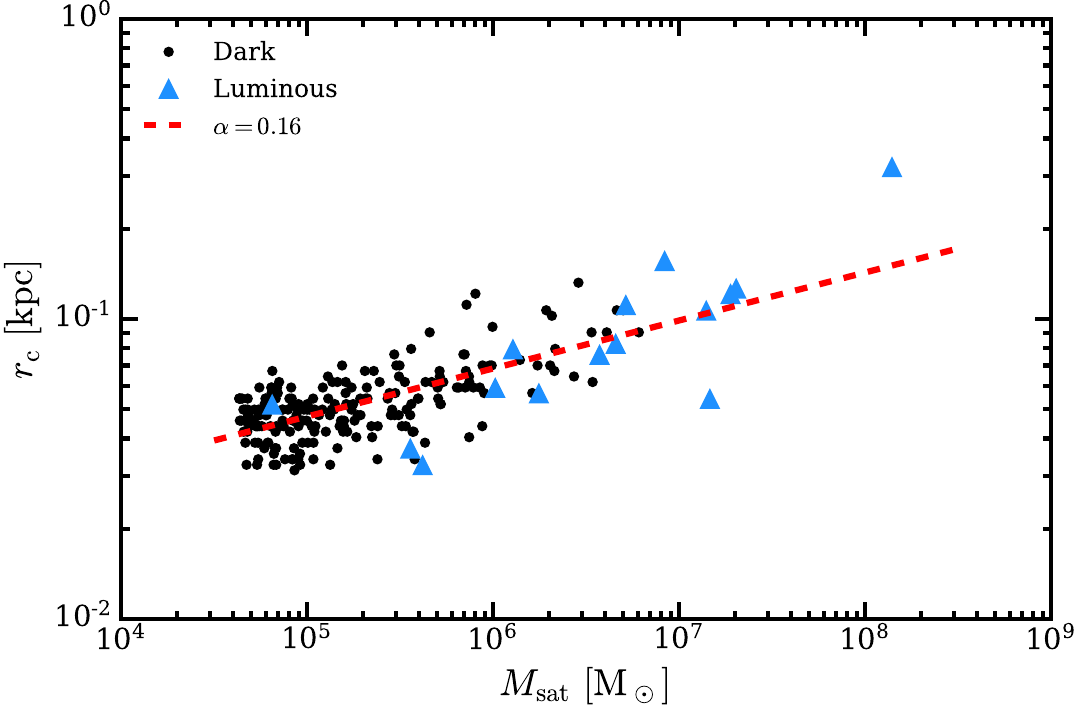}
    \includegraphics[width=0.49\textwidth]{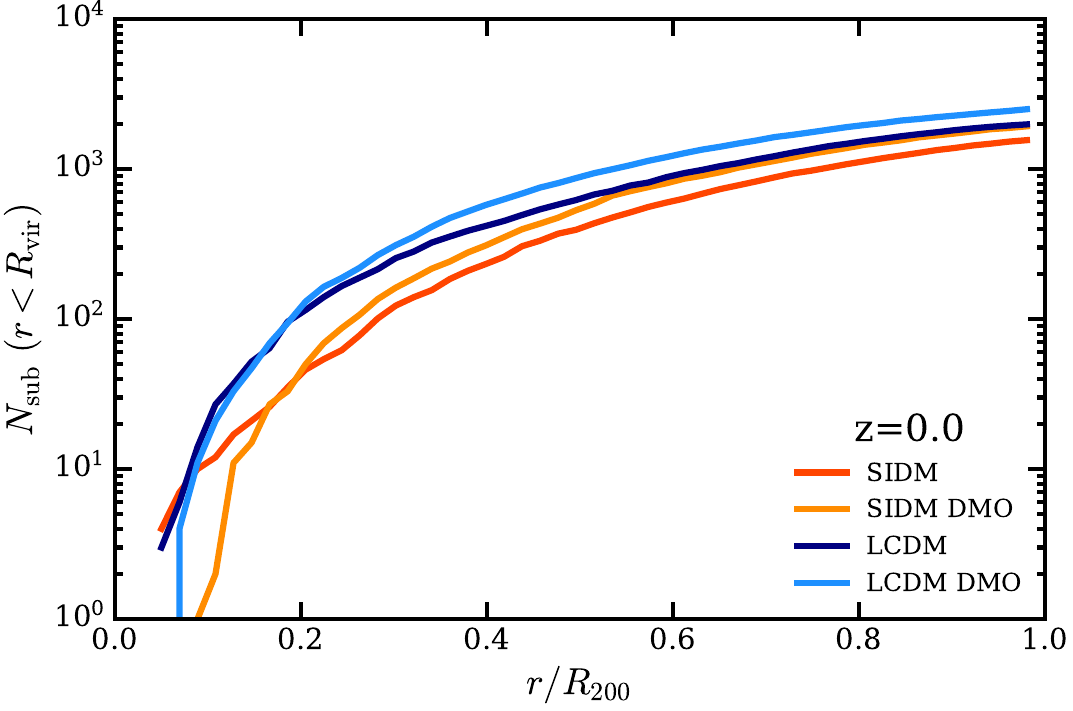}
    \caption{DM core size for satellites as a function of dynamical mass (left). Core sizes below 30 pc cannot be measured due to the gravitational softening of the DM. Right: The cumulative number of satellite halos more massive than $10^4\,\Msun$, as a function of the distance from the halo center (normalized by the virial radius, \Rvir). $\Lambda$CDM produces more centrally concentrated satellites, while the larger core in the SIDM model prevents satellites from falling to smaller radii. Cumulatively, the difference is 18\%, with 1984 and 2398 satellites within the virial radius, respectively.}
    \label{fig:coresize}
\end{figure*}

\subsection{Substructure}
\label{sec:subhalos}

There is both dark matter and stellar substructure within the main dwarf by $z=0$. We account for all the substructure in Tab.~\ref{tab:subhalos}. The substructure is detected in the simulation outputs with the post-processing algorithm \texttt{Subfind} \cite{Springel2001}. We distinguish three main categories: dark satellites, which are comprised solely of dark matter, luminous satellites comprised of dark matter and stars (and potentially some gas, although this is rare), and finally star clusters, defined as stellar-only systems. We further distinguish accreted and insitu-formed star clusters by checking whether the stars were inside the main halo at the time they formed. This distinction is
not made for the dark matter-containing subhalos, since all of them are accreted. Furthermore, we note that most accreted star clusters are hosted by a dark matter halo prior to being accreted. This halo is stripped away during infall and a stellar-only system remains \citep[see][for a detailed analysis]{Gutcke2024}. 

\begin{figure}
    \centering
    \includegraphics[width=\columnwidth]{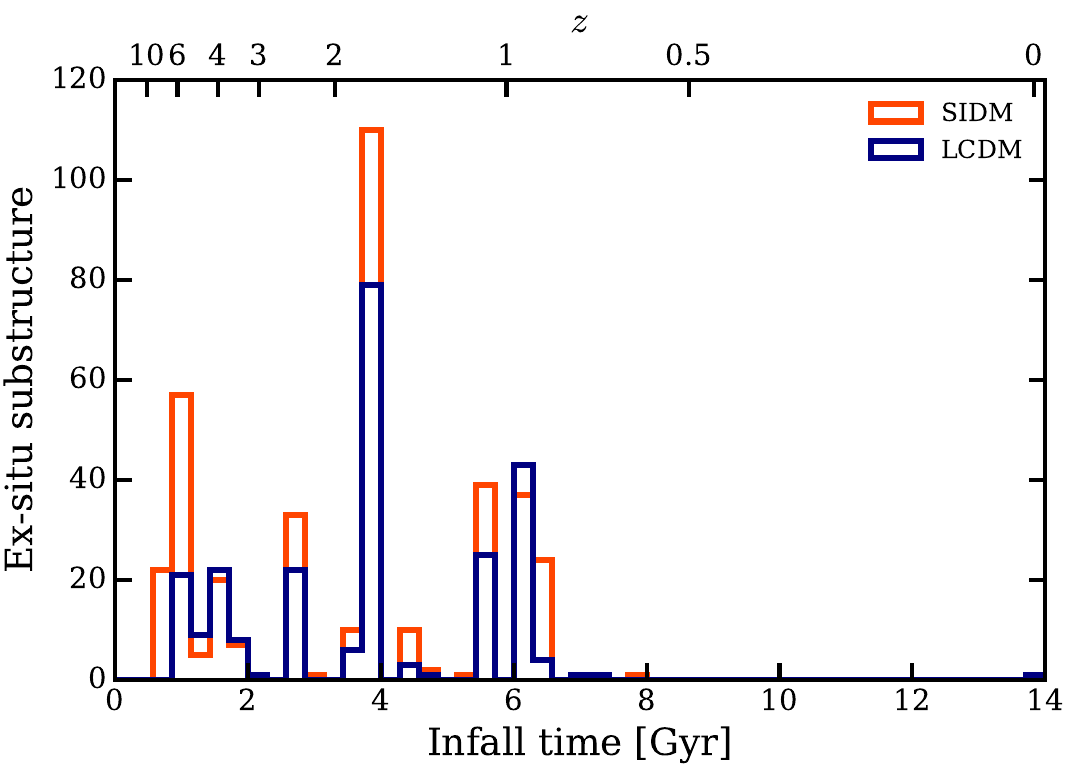}
    \includegraphics[width=\columnwidth]{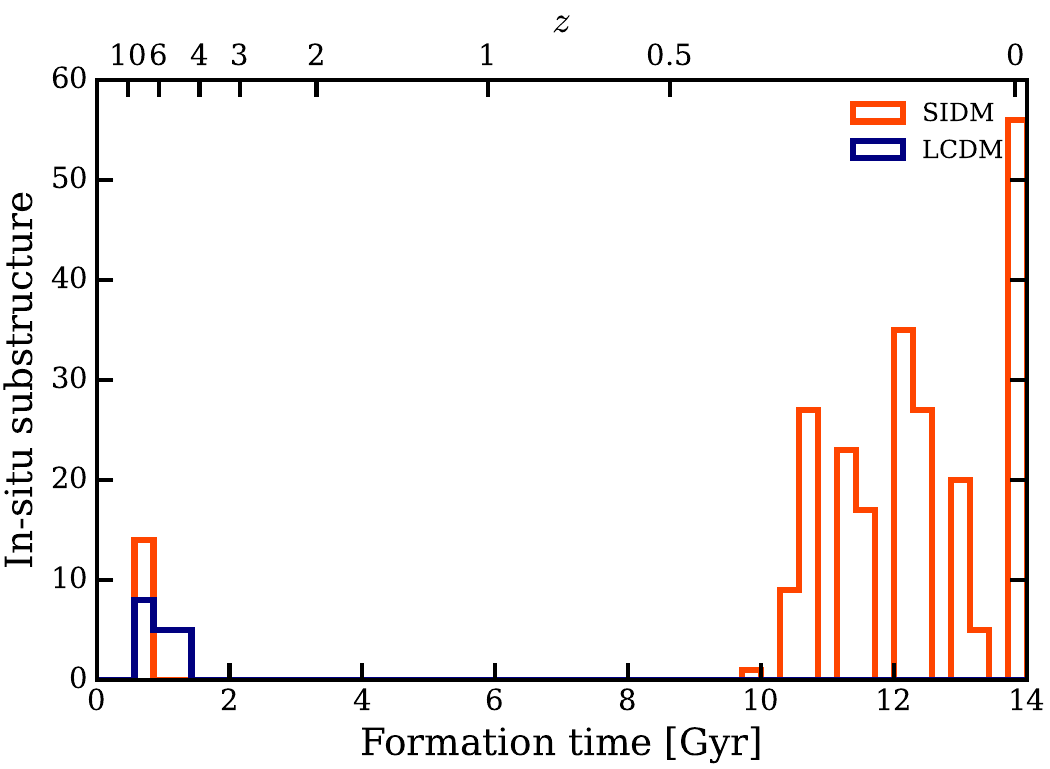}
    \caption{Top: Infall time of the ex-situ formed luminous substructure. Bottom: Formation time of the in-situ formed luminous substructure. All objects formed after $z=0.5$ are star clusters.}
    \label{fig:infall}
\end{figure}

As shown in Table~\ref{tab:subhalos}, the SIDM halo hosts roughly half as many luminous satellites as $\Lambda$CDM, but nearly five times more stellar-only star clusters. There is a stark distinction between the ex-situ substructure (objects formed a outside the main halo and later accreted) and the objects formed within the halo (in-situ formation). This is explored in Fig.~\ref{fig:infall} where we show the infall time into the main halo for the ex-situ objects (top) and the formation time for the in-situ objects in the lower panel. 
SIDM and $\Lambda$CDM mirror each other well in the times that subhalos are accreted, which occurs in the time range $z=10-1$. The impact of the SIDM cosmology is most apparent in the number of in-situ formed star clusters at late times, $z<0.5$. These have no counterpart in the $\Lambda$CDM cosmology and coincide directly with the re-onset of star formation (see Fig.~\ref{fig:sfh}). The extended DM core region in the SIDM galaxy allows the gas to remain distributed across a wider area. This in turn promotes the formation of separate star forming regions, which results in in-situ star clusters. This is different than in the $\Lambda$CDM case, where the majority of gas is funneled to the central cusp, forming a single, dense stellar body.

There are additional accreted satellites are stripped of their dark matter in the SIDM scenario, transforming into star clusters. It is interesting to consider the two distinct populations of star clusters found in the halo at the present day, namely ones that formed in the early universe ans were accreted and ones that formed within the halo at late times. This is reminiscent of the two populations of globular clusters \citep{Chen2023, Belokurov2024}.
We note that while core-collapse of SIDM subhalos can make them more resilient to stripping \citep{Penarrubia2010}, the subhalos that become star clusters here are too small to core-collapse (see Sec.~\ref{sec:corecollapse} below).

\begin{table}[]
    \centering
    \begin{tabular}{lllrrrr}
        &&& SIDM & \% & $\Lambda$CDM & \%\\
        \hline \hline
    \multicolumn{3}{l}{Total} &  6520 & & 7634 & \\
    & \multicolumn{2}{l}{Dark satellites}     &  5905  & 90.6 & 7368 & 96.5 \\
    & \multicolumn{2}{l}{Luminous satellites} &  55 & 0.8   & 126 & 1.7 \\
    & \multicolumn{2}{l}{Star clusters, no DM}  &  559 & 8.6  & 139 & 1.8 \\
    &&SCs accreted &  325  & 58.1  & 126 & 90.6 \\
    &&SCs formed insitu &  234  & 41.9  & 13 & 9.4 \\
    \hline \hline
    \end{tabular}
    \caption{Substructure counts identified with Subfind by category with percentages. Substructure is selected to be within the virial radius at $z=0$. Dark matter subhalos are included down to $M_\mathrm{DM} \geq 1.7\times10^3\Msun$. Star clusters are included down to $M_\star\geq 80\Msun$.}
    \label{tab:subhalos}
\end{table}


\subsection{Altered rotation curves and core-collapsed halos}
\label{sec:corecollapse}

\begin{figure*}
    \centering
    \includegraphics[width=\columnwidth]{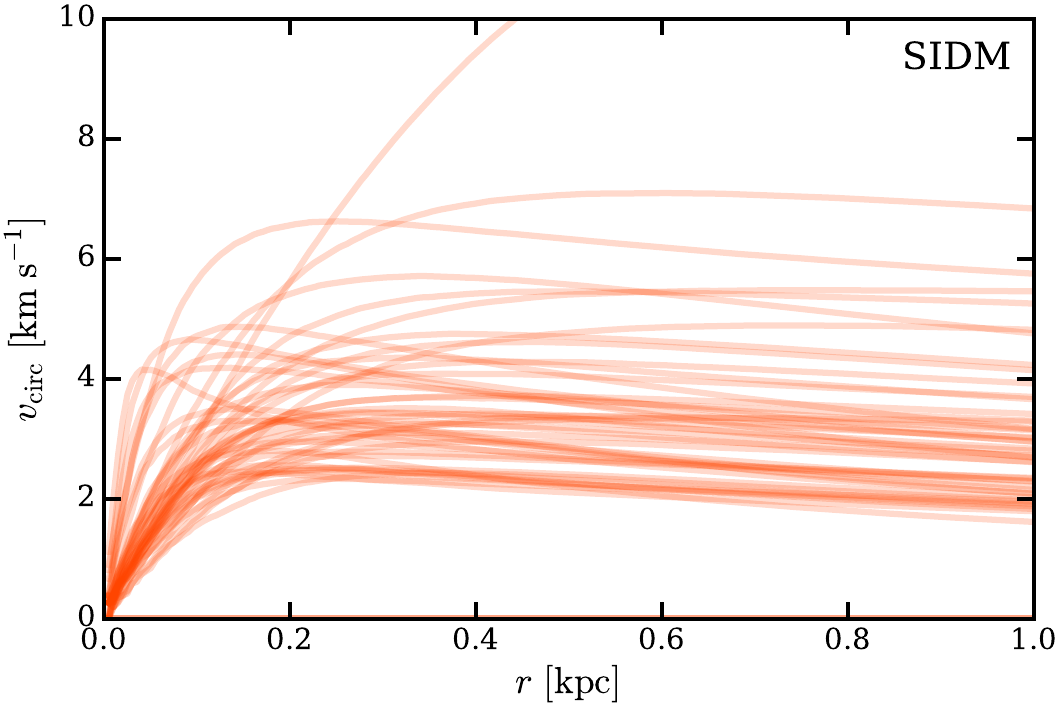}
    \includegraphics[width=\columnwidth]{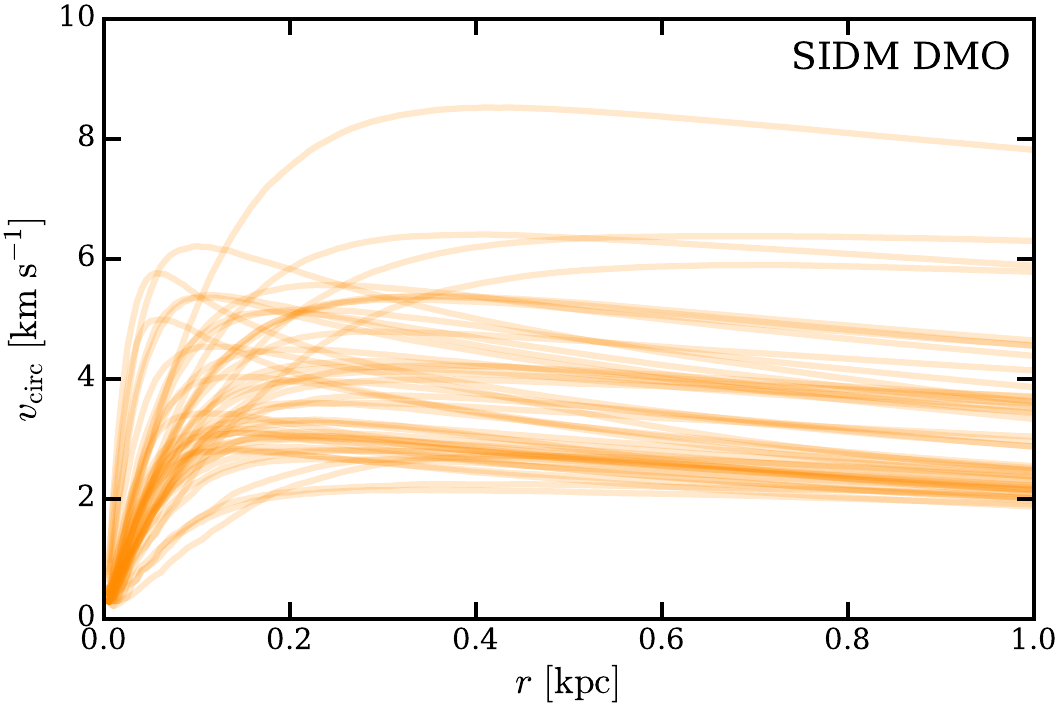}
    \includegraphics[width=\columnwidth]{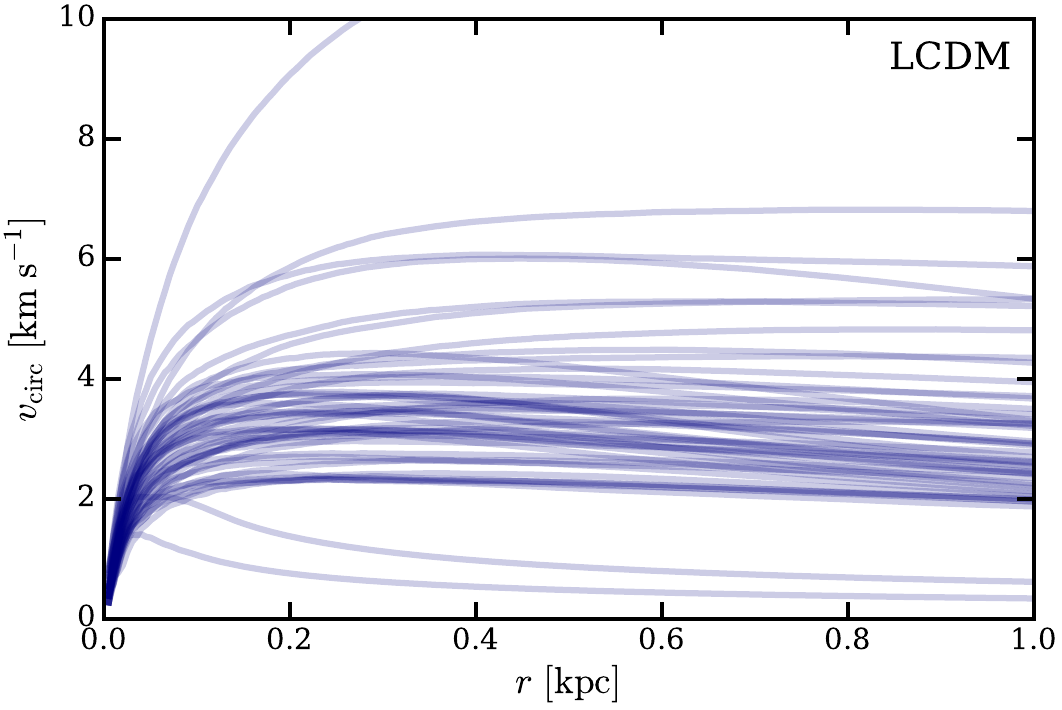}
    \includegraphics[width=\columnwidth]{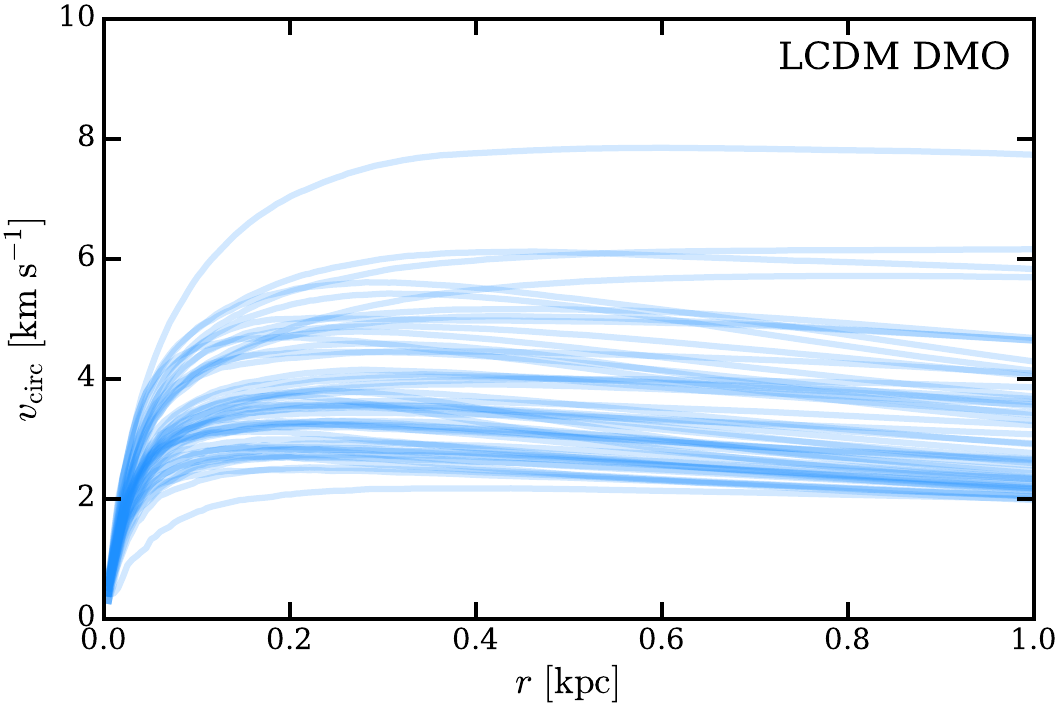}
    \caption{Rotation curves for the 50 most massive satellites in each galaxy. We can see there is more diversity of shapes in the two SIDM models.}
    \label{fig:rot}
\end{figure*}


A homogeneous density core is expected to contract and eventually undergo gravothermal collapse \citep{Balberg2002, Koda2011}. For isolated dark matter halos of galactic mass, this process typically takes longer than a Hubble time. However, if a dark matter halo enters the potential of a more massive host and experiences tidal stripping, some work has shown that the collapse may be significantly accelerated \citep{Nishikawa2020}.

It is not well known what will happen at scales below dwarf galaxies because this regime has not been studied in detail by simulations that are able to adequately resolve it with appropriate SIDM models. The expectation from previous models is that structures corresponding to the scale where the cross-section has its maximum and around the knee could undergo core-collapse, while those at higher masses and/or lower cross-section cannot.

\cite{Ando2025} recently presented the SASHIMI model, which provides a parametric mapping between CDM and SIDM subhalos that accurately reproduces SIDM density profiles and tidal evolution histories by capturing SIDM’s primary effects on the inner halo structure through the evolution of 
$v_\mathrm{max}$ and $r_\mathrm{max}$.
Predictions from this model show that the lowest masses do not reach core-collapse. The collapse is prevented by the concentration-mass relation. Most of the core-collapsed objects in their model are at intermediate scales.

As a caveat, most predictions to date are dark-matter-only models and it remains unknown how baryonic physics may change the predictions in the low-mass regime. If there is core-collapse then baryons are expected to accelerate it, but not many simulations have reached the necessary resolution to make firm predictions.

In Fig.~\ref{fig:rot} we present the rotation curves of the 50 most massive satellites. The title in each panel shows the simulation it presents. The lower panels ($\Lambda$CDM) show self-similar rotation curves that all rise and flatten in a similar fashion. The top panels, however, seems to show a more chaotic mixture of rotation curve shapes. The central rise of the various curves do not follow the same slope and the peaks also occur at various radii. This is encouraging, since the discrepancy between $\Lambda$CDM and observations is a lack of diversity of rotation curve shapes in $\Lambda$CDM models. 


We can take two measures of the shape of the rotation curve, namely $v_\mathrm{max}$, the peak of the curve, and $r_\mathrm{max}$, the radius at $v_\mathrm{max}$, to track core collapse more broadly. 

\begin{figure*}
    \centering
    \includegraphics[width=0.49\textwidth]{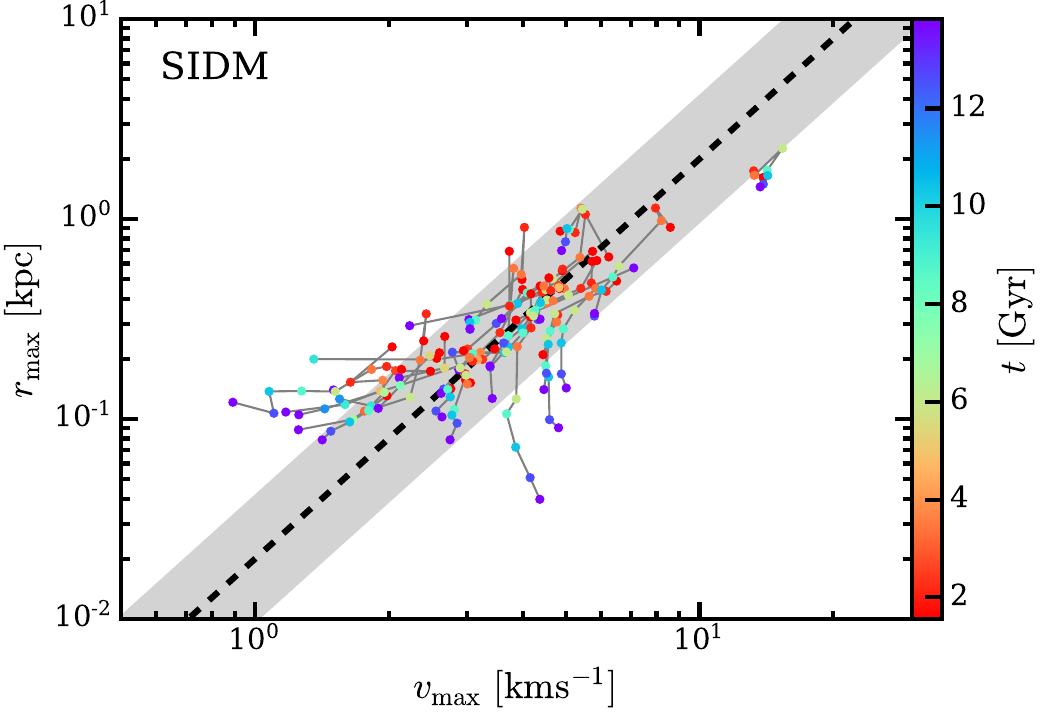}
    \includegraphics[width=0.49\textwidth]{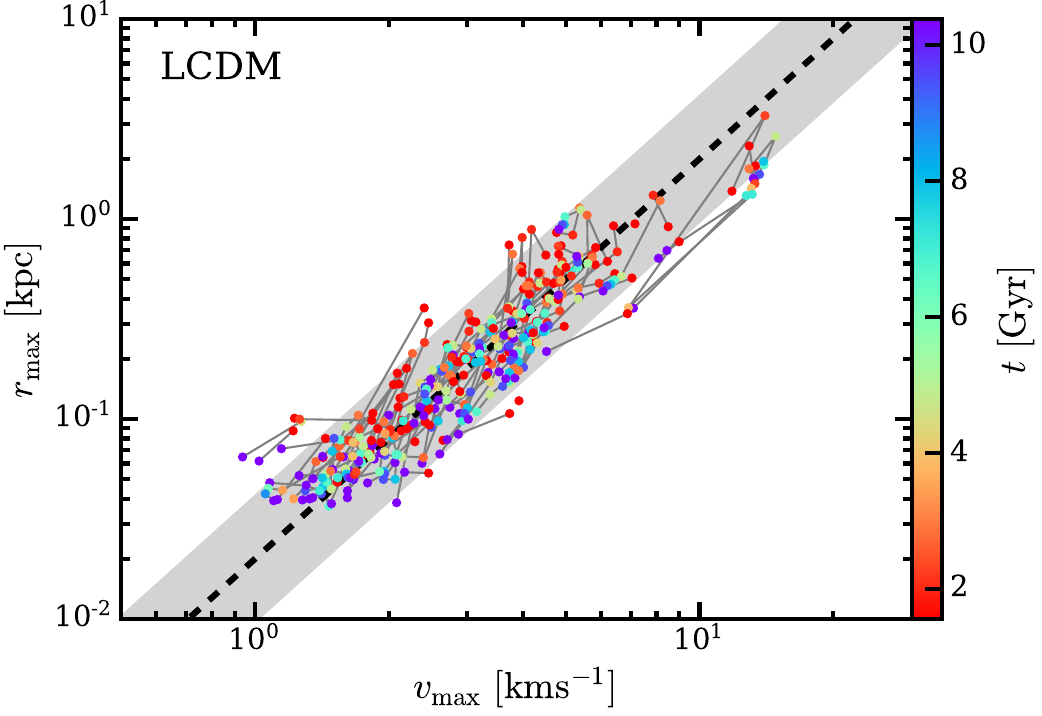}
    \caption{Evidence of core-collapse in satellites in SIDM (left) as opposed to $\Lambda$CDM (right). Figure shows evolutionary tracks of individual satellites on the $r_\mathrm{max}$ - $v_\mathrm{max}$ plane. The color of a dot along the track indicates the cosmic time as shown in the colorbar. Satellites were selected to have a resolved $r_\mathrm{max}$ at all times and to be dark matter dominated ($M_\mathrm{DM}>0$ and $N_\mathrm{DM}>N_\star$).}
    \label{fig:v-r_tracks}
\end{figure*}

In Fig.~\ref{fig:v-r_tracks}, we examine the $r_\mathrm{max}$–$v_\mathrm{max}$ plane to track the progression of core collapse in subhalos. Each gray line, connected by colored dots, represents the evolution of a single subhalo over time. The color of each dot indicates the cosmic time, as shown in the colorbar—red denotes early times, blue denotes later times. Motion down and to the right in this plane signals core collapse, corresponding to an increase in the peak circular velocity and a decrease in the radius at which it occurs. As seen in the comparison of the SIDM (left) and $\Lambda$CDM (right) panels, the rotation curves evolve differently in the two models: SIDM tracks generally move downward and slightly rightward, while $\Lambda$CDM tracks tend to move downward and to the left.

\begin{figure}
    \centering
    \includegraphics[width=0.49\textwidth]{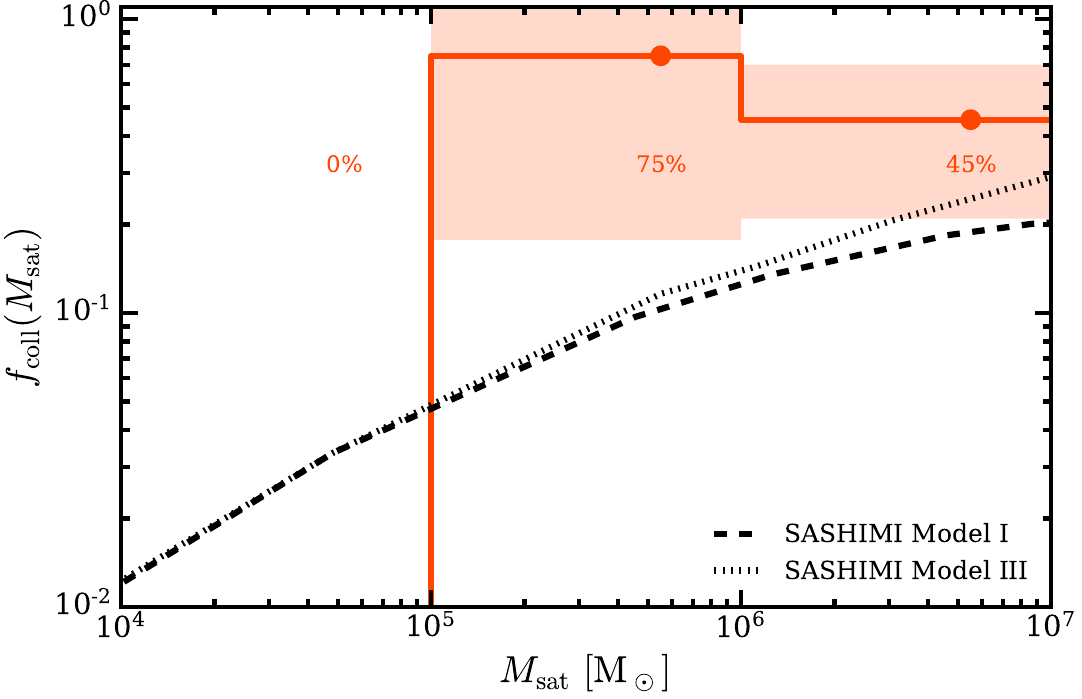}
    \caption{Fraction of collapsed satellites as a function of their dynamical mass in red with shaded area being the uncertainty range. Black lines are for comparsion with the SASHIMI semi-analytical model \citep{Ando2025}.}
    \label{fig:collapsed}
\end{figure}

We further quantify core collapse in the SIDM simulation by computing the collapsed fraction as a function of satellite dynamical mass (Fig.~\ref{fig:collapsed}). Subhalos are binned into three mass intervals: $(10^4, 10^5]$, $(10^5, 10^6]$, and $(10^6, 10^7]~M_\odot$. We compare our results to Models I and III from SASHIMI \citep{Ando2025}. In the highest mass bin ($10^6 < M/M_\odot < 10^7$), our simulation shows marginal agreement with Model III. However, in the intermediate mass bin, our hydrodynamical simulation predicts a significantly higher number of collapsed satellites. This discrepancy likely arises because the SASHIMI model does not include tidal stripping by the host halo, a process known to accelerate core collapse \citep{Sameie2018, Nishikawa2020}. The lowest mass bin includes only two satellites, neither of which are collapsed in our simulation. The low number of satellites in this bin stems from our selection criteria: these low-mass objects are often so heavily stripped of dark matter that they no longer satisfy our threshold of having more dark matter than stars. In some cases, stripping reduces them to the point that $r_\mathrm{max}$ is no longer resolved. Indeed, this lack of objects that satisfy the criteria is a direct result of the dark matter stripping that our subhalos undergo. Since the stellar orbits are much colder, they are not affected as strongly and remain as an increased number of star clusters in the galaxy.

In conclusion, our full hydrodynamical simulation predicts a higher collapsed fraction for subhalos with dynamical masses $M_\mathrm{dyn} > 10^5~M_\odot$ than models that do not account for spatial and orbital evolution within the host. Incorporating tidal stripping shifts the turnover mass of the collapsed fraction to lower values than previously predicted.


\section{Conclusions}
\label{sec:conclusion}

In this study, we compare the properties of a Local Group dwarf galaxy analogue evolved in two cosmological scenarios: the standard $\Lambda$CDM model and a self-interacting dark matter (SIDM) model with a velocity-dependent interaction cross-section. Both simulations are carried out using the LYRA model of galaxy formation \citep[][]{Gutcke2021} at high spatial and temporal resolution. Our analysis focuses on the global properties of the central dwarf galaxy as well as its substructure, including satellite galaxies and star clusters.

For the central dwarf, we find that neither SIDM nor baryonic physics significantly affect the overall growth history or final mass of the dark matter halo. Similarly, the velocity dispersion and rotation curve of the dwarf remain largely unchanged, consistent with previous findings \citep[e.g.,][]{Correa2025}. However, the SIDM model produces a cored dark matter density profile extending to approximately 1 kpc, about 2.8\% of the virial radius. Most of the dark matter displaced from the center to form the core remains in a slightly overdense shell surrounding the core (see the central panel of Fig.~\ref{fig:rhoDM}). Notably, the size of the core is relatively insensitive to the presence of baryons (top panel of Fig.~\ref{fig:rhoDM}).

Baryonic components, in contrast, respond sensitively to the SIDM-induced core. The total stellar mass of the dwarf is reduced by roughly 25\% in the SIDM simulation. A larger fraction of gas remains within the galaxy at $z=0$, with six times more gas mass enclosed within the stellar half-mass radius, compared to the $\Lambda$CDM counterpart. This reflects a markedly different star formation history: the SIDM model experiences an extended quiescent phase of about 8 Gyr between two distinct bursts at $z > 7$ and $z < 0.3$ (see Fig.~\ref{fig:sfh}). The altered distribution of baryons may yield observable differences: the stellar spheroid in SIDM is less centrally concentrated, and more stellar clusters form in-situ at late times (see Figs.~\ref{fig:barymap}, \ref{fig:starsigma} and \ref{fig:infall}).

We also investigate the substructure associated with the dwarf galaxy at $z=0$, including both satellite galaxies and stellar clusters. A striking difference emerges: fewer than half of the satellite dark matter halos that host stars in the $\Lambda$CDM simulation are luminous in the SIDM case. Moreover, satellites in $\Lambda$CDM tend to be more centrally concentrated than those in SIDM. The SIDM model also fosters a greater abundance and survival rate of stellar clusters; more than 40\% of them form in situ, compared to fewer than 10\% in $\Lambda$CDM (see Table~\ref{tab:subhalos}).

Interestingly, a subset of SIDM satellites undergo core collapse, but only after being accreted by the main dwarf, suggesting that the tidal field of the host triggers this process. Core collapse occurs in both hydrodynamical and dark matter-only runs, though all core-collapsed satellites are luminous in the full simulation. This leads to greater diversity in rotation curve shapes among SIDM satellites, with some but not all experiencing core collapse. While satellite core sizes scale with dynamical mass, they show no correlation with distance from the host.

Our analysis reveals that tidal stripping contributes to this behavior. The tidal field of the host halo removes outer dark matter layers, accelerating the collapse of the subhalo core.  This process can induce core collapse or transform satellites into stellar-only systems. Consequently, our hydrodynamical simulation predicts a higher collapsed fraction than semi-analytical models like SASHIMI, which do not account for the full orbital and spatial evolution of satellites within the host. This shifts the turnover mass for core collapse to lower values than previously anticipated, reinforcing the importance of environment and baryonic physics in shaping satellite structure.

\section*{Acknowledgments}
TAG acknowledges support by NASA through the NASA Hubble Fellowship grant $\#$HF2-51480 awarded by the Space Telescope Science Institute, which is operated by the Association of Universities for Research in Astronomy, Inc., for NASA, under contract NAS5-26555.
We acknowledge the computing time provided by the Leibniz Rechenzentrum (LRZ) of the Bayrische Akademie der Wissenschaften on the machine SuperMUC-NG (pn73we).
This research was also carried out on the High Performance Computing resources of the FREYA and COBRA clusters at the Max Planck Computing and Data Facility (MPCDF, \url{https://www.mpcdf.mpg.de}) in Garching operated by the Max Planck Society (MPG). GD acknowledges the funding by the European Union - NextGenerationEU, in the framework of the HPC project – “National Centre for HPC, Big Data and Quantum Computing” (PNRR - M4C2 - I1.4 - CN00000013 – CUP J33C22001170001). 


\bibliography{bib}{}
\bibliographystyle{aasjournal}


\end{document}